\documentclass[aps,prc,twocolumn,groupedaddress,showpacs]{revtex4}

\usepackage{graphicx}

\begin{document}
\title{
Further evidence for  a dynamically generated  secondary 
bow in $^{13}$C+$^{12}$C    rainbow scattering
 }

\author{S. Ohkubo$^{1}$,  Y. Hirabayashi$^2$
 and  A. A. Ogloblin$^3$,   
}
\affiliation{$^1$Research Center for Nuclear Physics, Osaka University, 
Ibaraki, Osaka 567-0047, Japan }
\affiliation{$^2$Information Initiative Center, Hokkaido University, Sapporo 060-0811, Japan}
\affiliation{$^3$RSC  ``Kurchatov Institute'',  RU-123182 Moscow, Russia}
 
\date{\today}

\begin{abstract}
 The existence of a secondary bow  is confirmed for  $^{13}$C+$^{12}$C nuclear rainbow
 scattering in addition to the $^{16}$O+$^{12}$C system. This is found by studying
 the experimental angular distribution of $^{13}$C+$^{12}$C
 scattering at the incident $^{13}$C energy $E_L$=250 MeV 
with an extended double folding 
(EDF) model that describes all the diagonal and off-diagonal coupling potentials 
derived from  the microscopic  wave functions for $^{12}$C  
  using  a density-dependent   nucleon-nucleon force.
The Airy minimum at $\theta$ $\approx$70$^\circ$, which is  not reproduced by a conventional
folding potential, is revealed to be a secondary bow generated  dynamically  by  a coupling
 to the excited state $2^+$ (4.44 MeV) of $^{12}$C.  
The essential importance of the quadruple {\it Y2} term (reorientation term) of  potential
 of the excited  state
 $2^+$ of $^{12}$C for the emergence of
 a secondary bow is found.
The mechanism of  the secondary bow is intuitively explained by showing 
how the  trajectories are refracted dynamically into the classically  forbidden 
angular region beyond the rainbow angle of  the primary rainbow. 
\end{abstract}

\maketitle

\par
Rainbows have been attracting mankind including poets and scientists 
\cite{Descartes,Newton,Nussenzveig1977,Greenler1980,Adam2002,Maitte2005}
 at least  two  thousand years.
A nuclear rainbow in the femto meter world   discovered by Goldberg
 {\it et al }\cite{Goldberg1974} is a Newton's zero-order rainbow \cite{Michel2002}, which was
 expected by Newton \cite{Newton} but not realized in   meteorological rainbow. 
The nuclear rainbows
 have  been extensively studied \cite{Khoa2007} and found to be very  important in the
 studies of   nuclear interactions  
\cite{Khoa2007,Khoa2000,Ogloblin2003,Ohkubo2004A,Ohkubo2007,Hamada2013,Ohkubo2014,Ohkubo2014B,Ohkubo2014C,Mackintosh2015,Ohkubo2015,Glukhov2007}
 and  nuclear cluster structures 
\cite{Michel1998,Ohkubo1999,Ohkubo2004B,Ohkubo2010,Hirabayashi2013}. 
The  nuclear rainbows, which   carry information  about 
 the deep inside of the  nucleus, can uniquely determine the interaction potential, 
that is,  the global potential which works over a wide range of energies from negative energy to
 the high energy region.
The existence of a secondary bow is not expected in principle in  a nuclear rainbow 
caused by refraction only.
In fact, in  the semiclassical theory of nuclear scattering 
\cite{Ford1959,Newton1966,Hodgson1978,Brink1985} in a 
 mean field nuclear potential, only one  extremum, (i.e., only one rainbow)  is allowed
 in the deflection function. 

\par
 Very recently the existence of a secondary bow  has been reported \cite{Ohkubo2014}
in the $^{16}$O+$^{12}$C rainbow  scattering at around $E_L$=300 MeV. 
Its existence 
 has not been  noticed in the conventional  optical model studies using a folding potential
 or a phenomelogical potential. 
The secondary bow  is  generated {\it dynamically} by a {\it quantum} coupling effect.
Via  quantum coupling to an excited state  of  $^{12}$C, a secondary bow emerges in the
 classically forbidden darkside of the ordinary (primary) rainbow caused by a mean field nuclear 
potential of a Luneburg lens \cite{Michel2002}. 
The  dynamical coupling has been shown to cause an additional attraction, which plays a role
 of a second lens, in the
 intermediate and inner region of the mean field Luneburg lens potential \cite{Mackintosh2015}.
It is  intriguing and important  to explore   whether a secondary rainbow, which is 
  logically not limited to the $^{16}$O+$^{12}$C system, is confirmed  in other systems.

\par
From this viewpoint, when we look carefully at the previously  observed  experimental data, 
 we notice that the rainbow
 scattering data for the $^{13}$C+$^{12}$C system available  at $E_L$($^{13}$C)=250 MeV 
\cite{Demyanova2010} show an  anomaly at large angles in the angular distribution  similar 
to  the  $^{16}$O+$^{12}$C system. It was not possible to describe it  
 in the  mean field optical potential model \cite{Demyanova2010}.

\par
 The purpose of this paper is to report the existence of a secondary bow in  
$^{13}$C+$^{12}$C scattering at $E_L$=250 MeV.
We investigate the angular distribution of  $^{13}$C+$^{12}$C rainbow scattering using 
the coupled channels (CC) method with an extended double  folding potential
 and show that a secondary bow is generated  dynamically   by the coupling to the $2^+$  
(4.44 MeV) state of $^{12}$C.  It is revealed that  
the quadruple {\it Y2} term (reorientation term) of  the 
 potential  
 of the  $2^+$ state   is essentially  responsible for  generating 
the  secondary  bow.

\par
We study   rainbow scattering for the $^{13}$C+$^{12}$C system with an extended double folding 
(EDF) model that describes all the diagonal and off-diagonal coupling potentials 
derived from  the microscopic   realistic wave functions for $^{12}$C  
  using  a density-dependent   nucleon-nucleon force.
  The diagonal and coupling potentials 
for the $^{13}$C+$^{12}$C system are calculated using the EDF  model as follows, 
\begin{eqnarray}
\lefteqn{V_{ij}({\bf R}) =
\int \rho_{00}^{\rm (^{13}C)} ({\bf r}_{1})\;
     \rho_{ij}^{\rm (^{12}C)} ({\bf r}_{2})} \nonumber\\
&& \times v_{\it NN} (E,\rho,{\bf r}_{1} + {\bf R} - {\bf r}_{2})\;
{\it d}{\bf r}_{1} {\it d}{\bf r}_{2} ,
\end{eqnarray}
\noindent where $\rho_{00}^{\rm (^{13}C)} ({\bf r})$ is the diagonal 
 nucleon  density of the ground state of  $^{13}$C   taken from Ref. \cite{DeVries1987}.
$\rho_{ij}^{\rm (^{12}C)} ({\bf r})$ represents the diagonal ($i=j$) or transition ($i\neq j$)
 nucleon density of $^{12}$C which is calculated using the microscopic three $\alpha$ cluster model 
in the resonating group method \cite{Kamimura1981}. This model reproduces the 
$\alpha$ cluster and shell-like structures of  $^{12}$C  well 
 and the  wave functions     have  been checked for many experimental
 data including charge form factors,  electric transition probabilities 
\cite{Kamimura1981} and the quadrupole moment of the $2^+$ (4.44 MeV) state \cite{Vermeer1983}.
We take into account the excitation of the  $2^+$  and $3^-$ (9.64 MeV) states
 of $^{12}$C in the calculations.    For the  effective interaction   $v_{\rm NN}$     we use  
 the DDM3Y-FR interaction \cite{Kobos1982}, which takes into account the
finite-range nucleon  exchange effect \cite{Khoa1994}.
We introduce the normalization factor  $N_R$ \cite{Satchler1979,Brandan1997,Khoa2001} for 
 the real double folding potential. 
An imaginary potential with a  Woods-Saxon volume-type (nondeformed) form factor is 
introduced   phenomenologically to take into account the effect
of absorption due to other channels.  It has been shown  in many coupled channel studies of 
 rainbow scattering involving  $^{12}$C 
\cite{Ohkubo2004A,Ohkubo2007,Hamada2013,Ohkubo2014,Ohkubo2014B} 
that  effects of densely populated  high-lying  excited  states,  including the energy region
 of    giant resonances, can be well expressed by an imaginary potential.
   A complex coupling, which is often used but has no rigorous theoretical justification
 especially when the projectile is  composite \cite{Satchler1983}, is not introduced because without it 
 the present EDF model  successfully reproduced  many  rainbow scattering data 
systematically over a wide range of incident energies 
\cite{Ohkubo2004A,Ohkubo2007,Hamada2013,Hirabayashi2013,Ohkubo2014,Ohkubo2014B,Ohkubo2014C,Mackintosh2015,Ohkubo2015} .

\begin{figure}  [t]
\includegraphics[keepaspectratio,width=8.7cm] {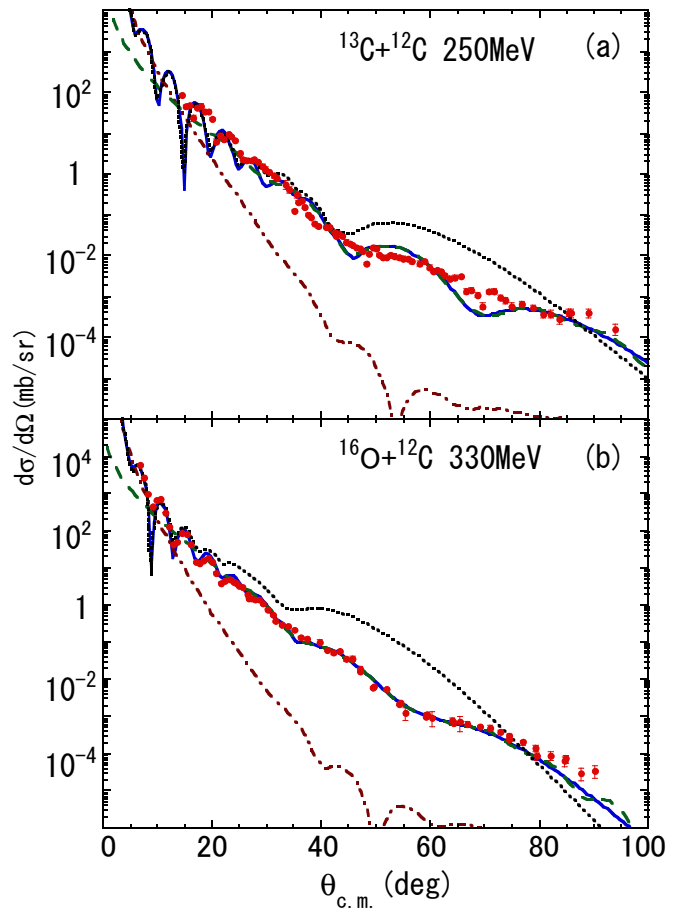}
 \protect\caption{\label{fig.1} {(Color online) 
   Angular distributions in (a) $^{13}$C+$^{12}$C  scattering at $E_L$=250 MeV
  and (b)  $^{16}$O+$^{12}$C scattering at $E_L=$330 MeV obtained using  the CC method 
(blue solid  lines) and single channel (dotted lines) calculations 
 are displayed in comparison with the experimental data (points) from 
Refs.\cite{Demyanova2010,Demyanova2004}.
 The farside and nearside components  of the CC calculations
are shown by the  green dashed lines  and the brown dash-dotted  lines, respectively.
The    CC  results are  obtained in panel (a)  with    coupling to the  $2^+$ 
and $3^-$ states of $^{12}$C and in panel (b) with   coupling to the $2^+$ and $3^-$ states
 of $^{12}$C and $^{16}$O. The potential parameters used in (b) are taken from 
Ref.\cite{Ohkubo2014}.}
}
\end{figure}

\par
 The nuclear rainbow  in  $^{13}$C+$^{12}$C scattering  was first observed by Bohlen {\it et al.}
 \cite{Bohlen1985} at $E_L$=260 MeV. The experimental angular distributions measured
  up to $\theta$=60$^\circ$ were reproduced  in the optical model  and CC
  calculations with a phenomenological Woods-Saxon potential. 
 Recently the measurement was extended at $E_L$=250 MeV to   larger angles up to 
$\theta$=94$^\circ$  in  Ref. \cite{Demyanova2010}.  
We note that it  was impossible  to reproduce the experimental angular distribution,
 which  does not  fall off monotonically beyond
  $\theta$$\approx$70$^\circ$, in the optical model
calculations with a phenomenological Woods-Saxon potential \cite{Demyanova2010}.  This
 discrepancy  between the calculation and the experimental data at  large angles
in rainbow scattering is reminiscent  of the similar difficult situation encountered in   $^{16}$O+$^{12}$C scattering
at around $E_L$=300 MeV \cite{Ogloblin2003}, which was solved by noticing the existence of a secondary bow generated dynamically
 \cite{Ohkubo2014}.

\par
 In Fig.~1(a)   the  angular distributions in elastic  $^{13}$C+$^{12}$C  scattering 
calculated  at $E_L$=250 MeV using the   single channel  double folding (DF)  potential  without
 channel  couplings  (dotted line)  and      CC with EDF  are 
displayed in comparison with the experimental data. 
We take $N_R$=1.2 for the real potential. For the imaginary potential the strength
 parameter  $W$=$20$ MeV  was found to 
fit the data  while  the radius parameter  and the diffuseness parameter 
 were fixed at   $R$=5.6 fm and   $a$=0.7 fm, respectively.  
The same    imaginary potential parameters are used both in the single channel
  and  CC calculations throughout this paper. 
We see that the single channel calculation gives the first Airy minimum $A1^{(P)}$ at $\theta$=45$^\circ$
 with the broad Airy maximum  $A1$ at $\theta$=55$^\circ$ followed by a falloff of the cross sections
in the darkside region. This $A1^{(P)}$ minimum corresponds well to  the observed Airy minimum at $\theta$=$45^\circ$ 
in the experimental data.  In the single channel calculation, however, the  structure
observed  beyond $\theta$=60$^\circ$ is missing.

\par
The CC calculation with coupling to the $2^+$  and $3^-$ states of $^{12}$C
is displayed by the blue solid line, which  does not fall off monotonically beyond 
$\theta$=60$^\circ$ and gives a minimum at $\theta$=70$^\circ$ in accordance with the experimental
 data. The calculated cross sections are decomposed into the farside (green dashed line)
 and nearside (brown dash-dotted line) components.  Beyond
 $\theta$=30$^\circ$ the nearside contribution decreases rapidly and
 the scattering is dominated by the refractive farside scattering. 
 It is difficult to see the difference between the  solid line and dashed line in Fig.~1(a).
 The minimum at $\theta$=70$^\circ$ is caused by refractive farside  scattering.
 Thus it is obvious that this minimum  located in the darkside 
region of  the primary rainbow is not caused by  the primary 
 rainbow due to the  Luneburg lens \cite{Michel2002} of the static mean field  nuclear potential. 
 It is the Airy minimum $A1^{(S)}$ of the secondary rainbow caused dynamically by the channel
 coupling to the excited states of $^{12}$C. This is reinforced by comparing it with 
the secondary  bow that appears in   $^{16}$O+$^{12}$C scattering systematically.  
 In Fig.1(b)  the angular distributions in
 elastic  $^{16}$O+$^{12}$C scattering at $E_L$=330 MeV are displayed.  
In the CC calculations with coupling to the $2^+_1$ and $3^-_1$ states of both $^{12}$C and 
$^{16}$O are included.
The behavior of the experimental and calculated  angular distributions in 
 $^{13}$C+$^{12}$C  scattering  resembles that of $^{16}$O+$^{12}$C  scattering
 where a secondary rainbow appears with the Airy minimum  $A1^{(S)}$ 
at around  $\theta$=$60^\circ$ in addition to the 
Airy minimum $A1^{(P)}$ of the primary rainbow at $\theta$=40$^\circ$ \cite{Ohkubo2014}.
We note here that  the contribution of the elastic transfer 
(one nucleon exchange)  contributions is small in the relevant angular region
 and does not contribute to the structure of the Airy minimum. In fact,
 Bohlen  {\it et al.} \cite{Bohlen1985} measured $^{12}$C($^{13}$C,$^{12}$C)$^{13}$C 
one nucleon  transfer reaction cross  sections  at $E_L$=260 MeV, which decrease rapidly toward 
large angles.

\begin{figure}[t]
\includegraphics[keepaspectratio,width=8.7cm] {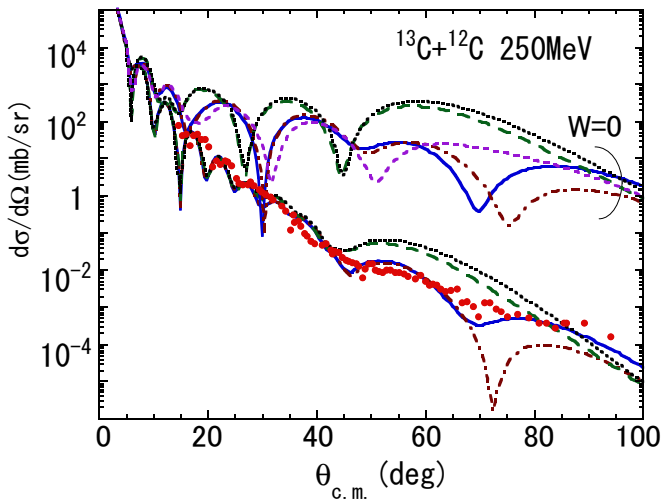}
\protect\caption{\label{fig.2} {(Color online) 
  Angular distributions in  $^{13}$C+$^{12}$C  scattering at $E_L$=250 MeV
 calculated using  the CC method  and a
single channel (black dotted lines)  
are displayed in comparison with the experimental data (points) \cite{Demyanova2010}.
 The blue solid, brown dash-dotted   and  green  long-dashed  lines
represent the CC results with    coupling to both  the $2^+$ 
and $3^-$ states of $^{12}$C,  those with  coupling to the $2^+$ state only and  those with
 coupling to the $3^-$ state only, respectively.
The upper figures ($W=0$) are calculated by switching off the imaginary potential.
The purple short-dashed line  represents the CC calculations with coupling to 
the $2^+$ state {\it without} the  quadruple {\it Y2} term (reorientation term) 
 for   $^{12}$C($2^+$). 
 }
}
\end{figure}

\par
We investigate what coupling is responsible for the generation of 
 the  Airy minimum $A1^{(S)}$ at $\theta$=70$^\circ$.
In Fig.~2 the  angular distributions calculated with coupling to the $2^+$ state only
and coupling to the $3^-$ state only are individually displayed.
 The calculated results with coupling to the $3^-$
 state (green dashed lines)  are essentially similar to those in the single channel calculation
(black dotted lines) and do not show the Airy minimum $A1^{(S)}$ 
at $\theta$=70$^\circ$.
 On the other hand,  the calculated results with coupling to the $2^+$ state 
(brown dash-dotted lines)  show  a deep Airy minimum at $\theta$=75$^\circ$.
 By including  the coupling to the $3^-$ state, as shown by the blue solid line, 
this deep Airy minimum is smeared and shifted
  forward slightly  approaching  the experimental Airy minimum. 
 The addition of
 coupling to the $3^-$ state  plays a role of introducing an imaginary 
potential. The origin of the Airy minimum is more clearly confirmed in the calculations by
 switching off the imaginary potentials.  The position of the Airy minimum in the calculation
 with coupling  to the $3^-$ state (W=0 green dashed line) is the same as that in the single
 channel calculation (W=0 black dotted line).  In the calculation with coupling to the $2^+$
 state with W=0 (brown dash-dotted lines), the additional Airy minimum is created 
at $\theta$=$75^\circ$, which is shifted forward
 about 5$^\circ$ by including the coupling to the $3^-$ state (blue solid). 
 It is clear that the
 Airy minimum  at $\theta$=$70^\circ$ is generated dynamically by the coupling to
 the $2^+$ state
 of $^{12}$C. The generation mechanism of the Airy minimum is completely different from that
at around $\theta$=40$^\circ$ due to the static mean field nuclear potential of
 a Luneburg lens \cite{Michel2002}.
Thus the Airy minimum $A1^{(S)}$ at $\theta$=70$^\circ$ is the same kind of  secondary 
bow that was  found 
in the   $^{16}$O+$^{12}$C  system  in Ref. \cite{Ohkubo2014}.

\par
We investigate further which part of  the quadruple {\it Y2} term (reorientation term) 
and  coupling potentials is  dominantly important in
 generating  the  secondary bow.
The diagonal potential  for  $^{12}$C($2^+$)  has  a $Y_2$ term  (reorientation term) 
 in the multipole expansion. 
  In Fig.~2   the  angular distribution 
 calculated using the CC method with coupling to 
the $2^+$ state but {\it without} the quadrupole $Y_2$ term  of the  
   potentials
  for   $^{12}$C($2^+$) is displayed for the 
  $W=0$ case by the purple short-dashed line. 
 The difference between  the brown dash-dotted line and the purple short-dashed line in the 
 $W=0$ case is due to   the quadruple {\it Y2} term  
 for the $2^+$ state. 
As we see  in the purple short-dashed line, the Airy minimum at $\theta$=75$^\circ$ of
 the secondary bow disappears  if the   $Y_2$ term  is not included.
 Thus it is found  that  the quadruple {\it Y2} term   of the 
 potential   for   $^{12}$C($2^+$) causes strong refraction 
 and  is essentially responsible
 for the  generation of   the  Airy minimum  of the secondary bow.

\begin{figure}[tbh]
\includegraphics[keepaspectratio,width=8.7cm] {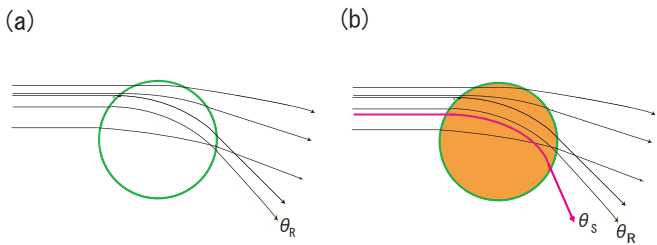}
 \protect\caption{\label{fig.3} {(Color online) 
The  illustrative  figures of  refractive   trajectories by the  attractive
 potential in nuclear rainbow scattering.
(a) Refraction in the case without coupling. The refracted angle $\theta_R$ is a rainbow angle
 for the  primary nuclear rainbow caused by the
 optical potential (Luneburg lens) of the nucleus (indicated by a circle). 
 The  angular region
 $\theta$$\leq $$\theta_R$ is the  bright side of the primary nuclear rainbow.
$\theta$$>$$\theta_R$ is not allowed  to refract classically and is the 
darkside.
(b) Refraction in the case  with coupling to the collective excited $2^+$ state of $^{12}$C.
 The  trajectories of a secondary bow (red line) strongly refracted (refracted angle 
$\theta_S$)  beyond $\theta_R$  by the  additional
    attractive potential   in the inner region induced    dynamically
 by the coupling  to the   excited state of the nucleus.
}
}
\end{figure}
 How a secondary bow is physically created by the coupling to the $^{12}$C($2^+$) 
was quantitatively investigated for the $^{16}$O+$^{12}$C system in Ref.\cite{Ohkubo2014}.
By using an inversion technique it was found in Ref.\cite{Mackintosh2015} that 
a dynamical  attractive potential is induced    by the
coupling to the collective excited state $2^+$ state  of $^{12}$C.
In Fig.~3  refraction  of the trajectories  in the classical picture 
in nuclear rainbow scattering is displayed.
 As shown in Fig.~3(a), when there is no coupling to the excited state,
 the refracted angle  $\theta_R$ is the largest (deflection) 
 angle (rainbow angle of the primary bow) among all the incident classical
 trajectories that are  refracted in the
nuclear optical potential (Luneburg lens). Refraction beyond $\theta_R$
 is classically forbidden.
 Therefore the angular region larger than $\theta_R$ becomes
 completely dark in classical mechanics. In quantum mechanics  cross sections 
falls rapidly beyond $\theta_R$   in the angular distribution as seen in the dotted line in 
 Fig.~1(a).
However, as shown in Fig.~3(b), when there is quantum coupling to the   $2^+$
 state of $^{12}$C state,
 classically forbidden refraction beyond $\theta_R$ becomes 
possible and  can be enhanced.  This is because the coupling to the collective excited 
state plays a role
 of a second lens, as discussed in Ref.\cite{Ohkubo2014}, creating a  
  characteristic polarization potential, especially
 attractive by nature in the internal region \cite{Mackintosh2015}.
As shown by the red line in Fig.~3(b), 
trajectories  can be  refracted  to larger angles beyond  $\theta_R$
by  the induced attraction in addition to the Luneburg lens potential. 
 Thus the refraction
 beyond  $\theta_R$ is enhanced  and the largest refractive angle, i.e., a second
 deflection angle $\theta_S$ appears  in the dark side of the primary nuclear rainbow. 
This is the secondary bow created on the darkside, $\theta>\theta_R$, 
 of the conventional primary nuclear rainbow as seen in the solid line in Fig.~1(a).

\par
Although there are no angular 
distribution  data available up to the large  angles around $\theta$=90$^\circ$ 
in  $^{13}$C+$^{12}$C elastic scattering except $E_L$=250 MeV,   it is theoretically expected
 that a secondary rainbow appears at other energies. 
In Fig.~4 the  angular distributions
 calculated using the   CC  method  with coupling to the  $2^+$ and $3^-$ states
 are displayed in comparison with  the single channel calculations for $E_L$=200, 260 and 330 MeV. 
In all the calculations $N_R$=1.2 for the real potentials and the same potential parameters
 are used  for the  imaginary
potentials. The real potential has energy dependence through that of 
 the two-body effective interaction DDM3Y-FR \cite{Kobos1982}. The volume integral per 
nucleon pair of 
the real potential, $J_V$, is 331, 316, 314, and  296  MeVfm$^3$ for  $E_L$=200, 250, 260 and
330 MeV, respectively. 
The calculations show  the emergence of  the Airy minimum $A1^{(S)}$ of the  secondary bow at
 $\theta\approx90^\circ$ for  $E_L$=200 MeV and at  $\theta\approx65^\circ$ for $E_L$=260 MeV.
At the higher energy of $E_L$=330 MeV  a sharper Airy minimum of the secondary bow is created
 dynamically by the channel coupling.  These Airy minima are  essentially created by the 
coupling to  the $2^+$ state of $^{12}$C.  As the incident energy increases, the Airy minimum
 $A1^{(S)}$  shifts to  forward  angles.
  It is  highly desired to measure the energy evolution of the Airy minimum in 
the angular distributions in $^{13}$C+$^{12}$C elastic scattering.
\begin{figure}[t]
\includegraphics[keepaspectratio,width=8.7cm] {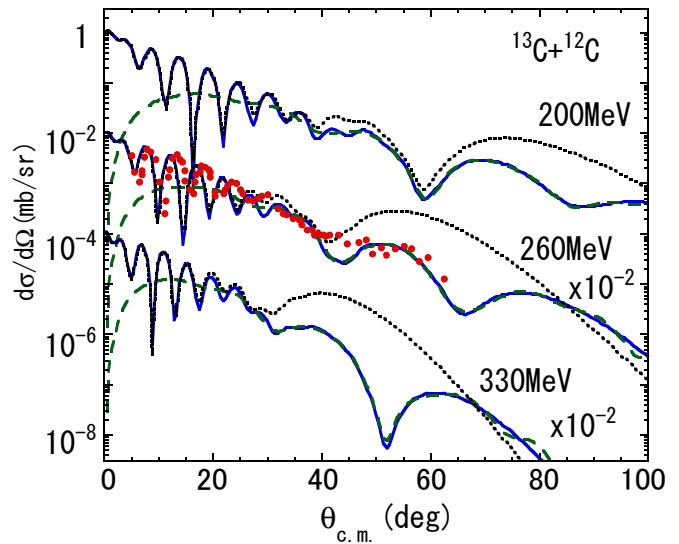}
 \protect\caption{\label{fig.4} {(Color online) 
 The energy evolution of the Airy structure in the angular distributions of the 
  cross sections in  $^{13}$C+$^{12}$C  scattering  calculated  with   coupling to the $2^+$ 
and $3^-$ states of $^{12}$C  (blue solid lines) and  the 
farside components (green dashed lines) are displayed  at $E_L$=200,  
260  and 330 MeV  in comparison with the experimental   data (points)
 from Ref.\cite{Bohlen1985}. 
The dotted  lines represent the  single channel calculation.  }
}
\end{figure}

\par
To summarize, we have shown the evidence for the existence of a secondary
  rainbow  in the angular distribution in   $^{13}$C+$^{12}$C scattering at $E_L$=250 MeV. 
 This was achieved by analyzing the experimental angular distribution using a coupled channel 
  method with an extended   double folding (EDF) potential derived from the microscopic   wave functions 
for $^{12}$C. 
The minimum at $\theta$$\approx$70$^\circ$ in the experimental angular distribution, 
which is not reproduced by the conventional
 optical potential model, is reproduced by the coupled channel calculations and found to be an
 Airy minimum of the secondary bow.
It is found that the secondary nuclear rainbow  is  caused by 
 coupling to the   $2^+$ state of $^{12}$C and  
the quadruple {\it Y2} term (reorientation term)  of the  potential 
  for the $2^+$ state  is essentially responsible for the creation of the secondary bow.
The mechanism of the generation of  the secondary bow is intuitively  explained by showing 
how the  trajectories are refracted dynamically into the classically  forbidden 
angular region beyond the rainbow angle of  the primary rainbow.

One of the authors (SO) thanks the Yukawa Institute for 
Theoretical Physics, Kyoto University  for  the hospitality extended  during a stay in May 2015.
A. O.  was partly supported by Russian Scientific Foundation 
(Grant No. RNF14-12-00079).

\end{document}